\documentclass[%
aip,
 amsmath,amssymb,
 reprint,%
]{revtex4-1}

\usepackage{graphicx}
\usepackage{dcolumn}
\usepackage{bm}

\usepackage[utf8]{inputenc}
\usepackage[T1]{fontenc}
\usepackage{mathptmx}

\def\ton{\tilde{t}^{\rm (on)}}
\def\toff{\tilde{t}^{\rm (off)}}
\def\jj{{\rm j}}
\def\dd{{\rm d}}
\def\U#1{\,{\rm #1}}
\def\vo2{{$\rm VO_2$}}

\begin{document}

\title[]{
Broadband operation of active terahertz quarter-wave plate achieved with
vanadium-dioxide-based metasurface switchable by current injection
\vspace{3mm}}

\author{Toshihiro Nakanishi}
 \affiliation{Department of Electronic Science and Engineering, Kyoto University, Kyoto 615-8510, Japan}
 \email{t-naka@kuee.kyoto-u.ac.jp}
\author{Yosuke Nakata}%
 \affiliation{Graduate School of Engineering Science, Osaka University, Osaka 560-8531, Japan}
\author{Yoshiro Urade}
\affiliation{%
Center for Emergent Matter Science, RIKEN, Saitama 351-0198, Japan}%
\author{Kunio Okimura}
\affiliation{
School of Engineering, Tokai University, Kanagawa 259-1292, Japan}

 \date{\today}

\begin{abstract}
 We demonstrate the broadband operation of a switchable terahertz quarter-wave plate
 achieved with an active metasurface employing vanadium dioxide.
 For this purpose, we utilize anisotropically deformed checkerboard structures,
 which present broadband characteristics compatible with deep modulation.
 Moreover, the metasurface is integrated with a
 current injection circuit to achieve state switching;
 this injection circuit can also be employed 
 to monitor the electric state of vanadium dioxide.
 We estimate the Stokes parameters derived from the experimental
 transmission spectra of the fabricated metasurface
 and confirm the helicity switching of circularly polarized waves
 near a designed frequency of $0.66 \U{THz}$.
 The relative bandwidth is evaluated as 0.52,
 which is 4.2 times broader than that in a previous study.
\end{abstract}

\maketitle

Terahertz waves have unique properties, such as 
high transparency to optically opaque materials,
and distinct spectral responses to molecules;
they have been extensively applied in
nondestructive imaging,
biomaterial detection,
and other areas \cite{Yun-Shik,Peiponen}.
In the development of terahertz technologies,
it is quite important to manipulate polarization,
which is one of the most fundamental characteristics of
electromagnetic waves.
In optical regions, a birefringent material is used to realize
polarization devices, such as a half- or quarter-wave plate; 
however, the available terahertz components for polarization control
are still limited and inefficient.
Furthermore, active control of terahertz polarization 
is much more difficult.

Recently, artificial materials composed of designed subwavelength
structures, called as metamaterials, have gained significant interest
for the manipulation of electromagnetic waves\cite {Solymar}.
Numerous investigations have focused on
two-dimensional subwavelength structures, called as metasurfaces\cite{Chen2016},
which induce discontinuous phase shifts in transmitted or reflected
electromagnetic waves\cite{Yu2011, Pfeiffer2013}.
Generally, anisotropic and chiral metasurfaces
can modify the polarization of electromagnetic waves, owing to the phase difference between orthogonal
polarizations, and various types of metasurfaces have been proposed
to provide a birefringent response \cite{Pfeiffer2014a} or optical activity \cite{Rogacheva2006}.
In addition,
active polarization control can be realized by reconfigurable metasurfaces
incorporating dynamic elements,
such as microelectromechanical systems (MEMS)\cite{Kan2013, Zhang2017,
Zhao2018, Cong2017}, semiconductors
\cite{Zhang2012b}, and graphene \cite{Miao2015}.
Phase-change materials, such as vanadium dioxide (${\rm VO_2}$)
(which undergoes an insulator-to-metal transition near $65 \, ^\circ{\rm C}$),
 have also been utilized
to achieve active polarization control \cite{Wang2015, Wang2016a,
Liu2018b, Nouman2018}.
Previously, we have also proposed a \vo2-integrated metasurface
with dipole-embedded checkerboard structures functioning as an
active quarter-wave plate, whose fast and slow axes can be interchanged
by increasing the temperature using an external
heater\cite{Nakata2018}.
This metasurface can reverse
the rotational direction of circularly polarized waves
that are generated from linearly polarized incident waves.
Generally, the broadband operation of metasurfaces is challenging, because
they frequently employ resonances to induce effective electromagnetic responses.
The above-mentioned active terahertz quarter-wave plate also suffers from a severely limited bandwidth,
owing to its complex spectral response unique to Fano
resonances\cite{Miroshnichenko2010,Kamenetskii}.
This is resulting from the interference between the broad resonance of the
checkerboard structures and the sharp resonance of the dipole
structures that are embedded in the checkerboard structures to
induce an anisotropic response.

In this study, we significantly broaden 
the operation bandwidth of a metasurface
functioning as an active quarter-wave plate.
For this purpose, we take advantage of the broadband responses inherent to
checkerboard structures \cite{Takano2014, Urade2016},
without the use of dipole structures,
which adversely affect these responses.
Instead of introducing dipole structures,
we anisotropically deform the checkerboard structures and
achieve broadband operation of the active quarter-wave plate.
In conjunction with bandwidth broadening, 
we integrate a current injection circuit with the metasurface,
to induce a phase transition.
The state of the metasurface can be 
controlled by injecting an electric current into the
\vo2 sheets incorporated in it,
and the electric state of \vo2
is identified by monitoring the injecting current
and the applied voltage.
This integrated design of a metasurface without an external heater
is suitable for the miniaturization of a device.

\begin{figure}[t]
 \begin{center}
  \includegraphics[]{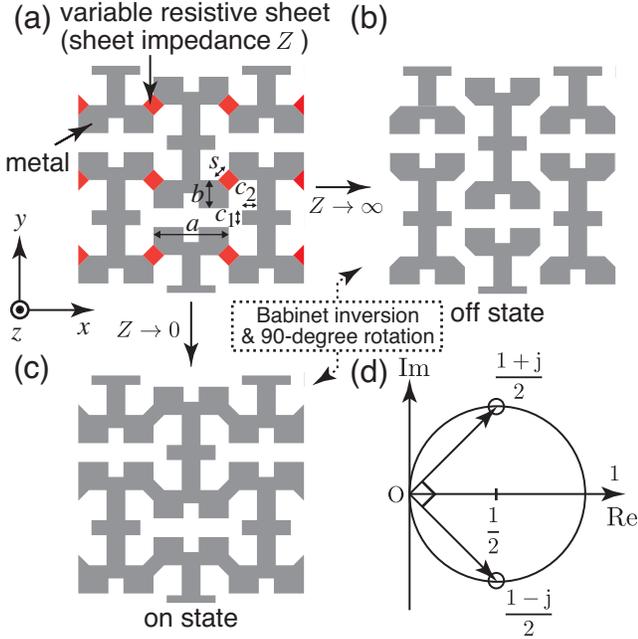}
  \caption{(a) Top view of schematic design of metasurface. (b) Off state ($Z\rightarrow \infty$). (c) On state ($Z \rightarrow 0$).
(d) Amplitude transmission coefficients plotted on a complex plane for realizing quarter-wave plate.}
  \label{design}
 \end{center}
\end{figure}

We briefly review the design rules based on Babinet's principle\cite{Nakata2018}
using an actual metasurface,
whose top view is shown in Fig.~\ref{design}(a).
The metasurface is composed of metallic sheets and variable resistive sheets whose sheet impedance $Z$ can vary over a wide range.
Figures~\ref{design}(b) and (c) illustrate the states, labeled as 
off and on states,
in the limit of $Z \rightarrow \infty$ and $Z \rightarrow 0$, respectively.
These two states are complementary to each other, for the inversion of 
the metallic and vacant parts.
The on state can be obtained by rotating the inverted structure of the off state
by 90 degrees, and vice versa.
From Babinet's principle applying to the metasurface with the distinct
symmetry,
the following relations can be derived for the complex amplitude transmission coefficients:
\begin{equation}
 \toff_x + \ton_x = 1,\quad  \toff_y + \ton_y=1, \label{Babinet}
\end{equation}
where subscripts $x$ and $y$ represent the corresponding polarizations of the
incident waves,
and superscripts (on) and (off) represent the states of the variable
resistive sheets\cite{Nakata2013,Nakata2016a}.
If the single-layer metasurface with a subwavelength thickness 
does not have any loss, including the Ohmic dissipation and energy leakage
resulting from diffraction
and polarization conversion between the $x$ and $y$ polarizations,
the transmission coefficients should be located on the circumference of a 
unit circle with the center at $1/2$ in a complex plane, as depicted in
Fig.~\ref{design}(d).
Assuming that the metasurface in the off state is designed to act as a
quarter-wave plate, which demands $\toff_x = \pm \jj \, \toff_y$,
possible solutions are provided as  $\toff_x =(1 \pm \jj)/2$ and $\toff_y=(1 \mp \jj)/2$, respectively.
They are shown as open circles in Fig.~\ref{design}(d).
In this case, the magnitudes of the transmission coefficients should satisfy
the following conditions\cite{Nakata2018}:
\begin{equation}
 |\toff_x(\omega)| = |\toff_y(\omega)|, \quad \frac{\dd |\toff_x|}{\dd \omega} \cdot \frac{\dd |\toff_y|}{\dd \omega}<0,
 \label{condition}
\end{equation}
where $\omega$ is an angular frequency.
Babinet's relations, as expressed in Eq.~(\ref{Babinet}), ensure that
the on state also functions as a quarter-wave plate with $\ton_x = (1 \mp \jj)/2 \,(=\toff_y)$
 and $\ton_y = (1 \pm \jj)/2 \,(= \toff_x)$,
where the slow and fast axes are interchanged compared to those for the off state.

\begin{figure}[t]
 \begin{center}
  \includegraphics[]{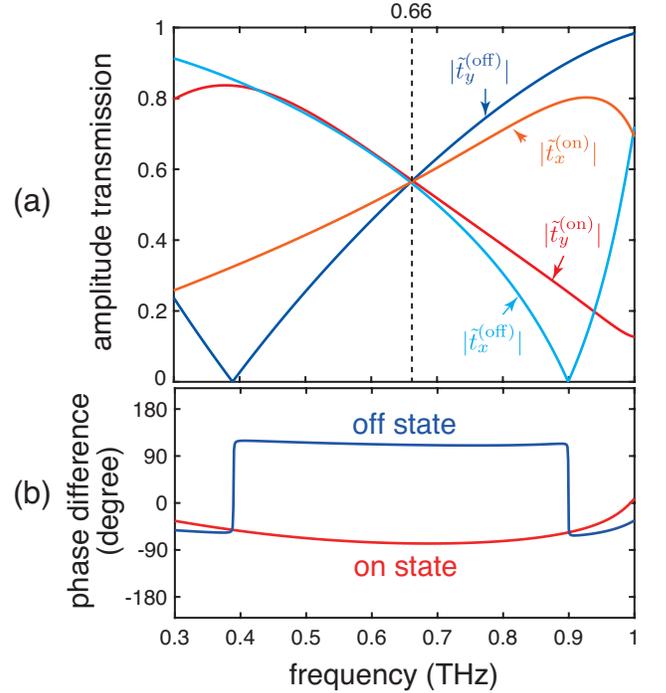}
  \caption{(a) Normalized amplitude transmission spectra and (b) phase differences between $y$ and $x$ polarization components derived from simulation results.}
  \label{simulation2}
 \end{center}
\end{figure}

Based on the above strategy, 
we design a metasurface by adjusting the dimensions of the structures
as shown in Fig.~\ref{design}(a),
using a commercial software package (CST Microwave Studio).
We suppose that the metallic sheets are composed of perfect electric conductors 
with zero thickness and that the variable resistive sheets are in 
an insulating state with $Z = 100 \U{k\Omega}$,
which is a typical sheet impedance of a 200-nm-thick VO$_2$ film at room temperature $\sim 25\U{^\circ C}$ (see in Supplementary Material).
The metasurface is formed on a {\it c}-cut sapphire substrate with
semi-infinite thickness and 
anisotropic refractive indices of $n_x = n_y = 3.1$ in the $x$--$y$ plane and $n_z = 3.4$ in the propagation direction\cite{Grischkowsky1990}. 
We fixed $s$ and $a$ as $s = 10 \U{\mu m}$ and $a = 60 \U{\mu m}$, respectively,
which determine diffraction frequency $f = c_0/(\sqrt{2} n_z a)$ near $1.04 \U{THz}$.
Amplitude transmission spectra, which are normalized by that obtained
for the substrate without
the metasurface, for normally incident terahertz waves
are calculated for the periodic system in the $x$ and $y$
directions under the periodic boundary conditions.
From the simulation results, we determine the design parameters as
$b = 25 \U{\mu m}$, $c_1 = 14 \U{\mu m}$, and $c_2 = 15 \U{\mu m}$,
such that Eq.~(\ref{condition}) for the off state is satisfied.
The details of the optimization procedure are described in the Supplementary Material.
Figure~\ref{simulation2}(a) presents the calculated transmission spectra.
The sheet impedance of the resistive sheets for the on state is set as $Z = 10 \U{\Omega}$.
As expected, for the on state, two transmission spectra cross at the intersection point for the off state
near $0.66 \U{THz}$,
and Eq.~(\ref{condition}) is also automatically satisfied for the on state.
Figure~\ref{simulation2}(b) shows the phase difference between the $y$ and $x$ polarization components, defined as
 ${\rm arg}(\tilde{t}_y/\tilde{t}_x)$, for both the states.
The practically flat response found in the broad spectral range from $0.4 \U{THz}$ to $0.9 \U{THz}$ 
is a unique property of this metasurface.
The phase differences at $0.66 
\U{THz}$ are estimated as $+111\U{^\circ}$ and $-78\U{^\circ}$
for the off and on states, respectively,
and they slightly deviate from the ideal values of $\pm 90\U{^\circ}$, respectively.
This is because Eqs.~(\ref{Babinet}) and (\ref{condition})
are not strictly satisfied, mainly because the substrate breaks the reflection symmetry required for Babinet's principle.  
Nevertheless, the metasurface presents excellent performance as a linear-to-circular polarization converter,
which will be discussed below using  experimental results.

In our experimental demonstration, variable resistive sheets are
composed of \vo2,
which exhibits insulator-to-metal transitions above the critical temperature of approximately $65\U{^\circ C}$. 
Figure~\ref{photo}(a) shows a photomicrograph of the 
metasurface fabricated on a {\it c}-cut sapphire substrate.
The thicknesses of the \vo2 and aluminum films are estimated as approximately
$200 \U{nm}$ and $400 \U{nm}$, respectively.
The whole structure is presented in Fig.~\ref{photo}(b).
The \vo2 patterns are formed by wet etching for the \vo2 film deposited by reactive magnetron sputtering, and metallic patterns are formed by lift-off process \cite{Urade2016}.
The details of the fabrication procedure
are provided in the Supplementary Material.
A metasurface with a size of $12 \U{mm} \times 9 \U{mm}$ is fabricated at the center of the substrate.
For monitoring and controlling the electric state of the \vo2 films,
two electrodes are introduced at the top and bottom ends of the metasurface,
to inject electric currents into the \vo2 patches.
The electric currents are applied through electric wires,
which are connected to each electrode with a conductive adhesive.
Both left and right sides of the metasurface are covered by $24\U{nm}$-thick-titanium films with a width of $1.5\U{mm}$.
The role of the titanium films is discussed subsequently.

\begin{figure}[]
 \begin{center}
  \includegraphics[]{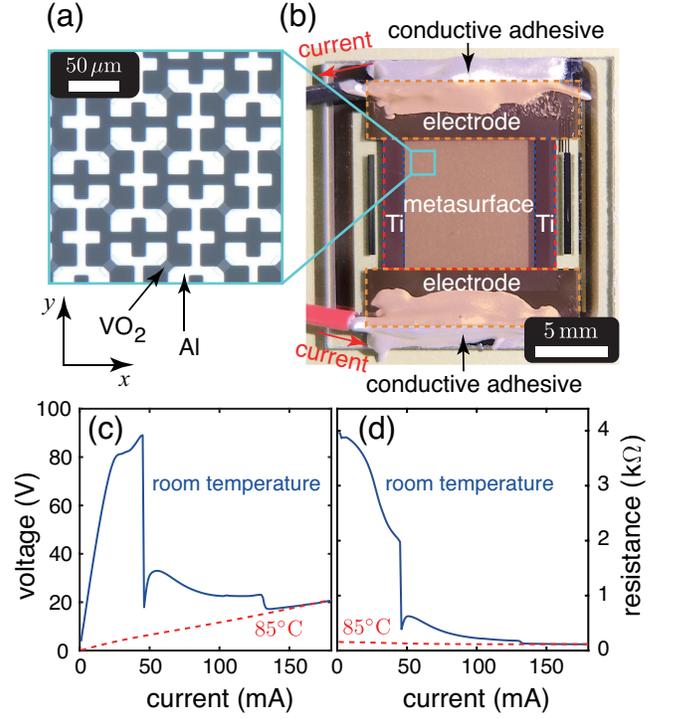}
  \caption{(a) Photomicrograph of metasurface. (b) Whole structure of device. (c) Current--voltage characteristics between electrodes. (d) Current--resistance characteristics between electrodes.}
  \label{photo}
 \end{center}
 \end{figure}
 
Before terahertz measurements,
the current--voltage (I--V) characteristics between the electrodes are evaluated
using a direct-current power source operated in a constant-current mode.
The results observed at room temperature around $24\U{^\circ C}$ with
increasing current at a rate of $0.5\U{mA/s}$ are
shown as a solid line in Fig.~\ref{photo}(c).
The dashed line represents the I--V characteristics
when the mount holding the metasurface is heated at $85\U{^\circ C}$.
The resistance characteristics $R=V/I$ derived from the I--V characteristics are shown 
in Fig.~\ref{photo}(d). For small current $I<20\U{mA}$, 
the voltage increases almost linearly with increasing current.
In this region, most of the electric current is concentrated on the titanium films at the sides of the metasurface, and the resistance $R\sim 4\U{k\Omega}$
can be regarded as the resistance of the titanium films.
This is because the sheet impedance of the \vo2 films is extremely high
($\sim 67 \U{k\Omega}$)
at room temperature.
The Joule heat in the titanium sheet increases the temperature of the sapphire substrate with a high 
heat conductance and facilitates the phase transitions of the \vo2 films.
At approximately $I = 45\U{mA}$,
the voltage across the electrodes abruptly drops from $V = 90 \U{V}$,
which suggests that some of the \vo2 patches undergo a phase transition,
and  conducting paths in the metasurface are formed.
The required voltage for the phase transition would be considerably higher without the titanium sheets,
which effectively reduce the threshold voltage. 
With increasing $I$, a ratio of \vo2 patches in the metallic state is gradually increased. 
The I--V characteristics present a small drop close to $I = 130 \U{mA}$
and asymptotically approach those of the dashed line.
This suggests that the \vo2 patches are completely in the metallic state for $I > 130\U{mA}$,
because the dashed line represents the I--V curve at a temperature considerably higher than the critical temperature.
The resistance approaches to a constant value around $120\U{\Omega}$, 
which is much smaller than the resistance of the titanium films $\sim 4\U{k\Omega}$,
and most of the current is concentrated on the metasurface without the titanium films.
The two-step transition in Fig.~\ref{photo}(c)
is also observed for the conduction of a single \vo2 gap,
owing to the percolation processes associated with the metallic and
insulating phases coexisting in a metastable state\cite{Zhao2012}.
The state at $I = 180 \U{mA}$ ($V = 21 \U{V}$), which corresponds to a power consumption of $3.8\U{W}$, was used 
as the on state in the following experiments.
 The power consumption in the titanium films is estimated to be $0.11\U{W}$, which is three percent of the total power consumption.

\begin{figure}[]
 \begin{center}
  \includegraphics[]{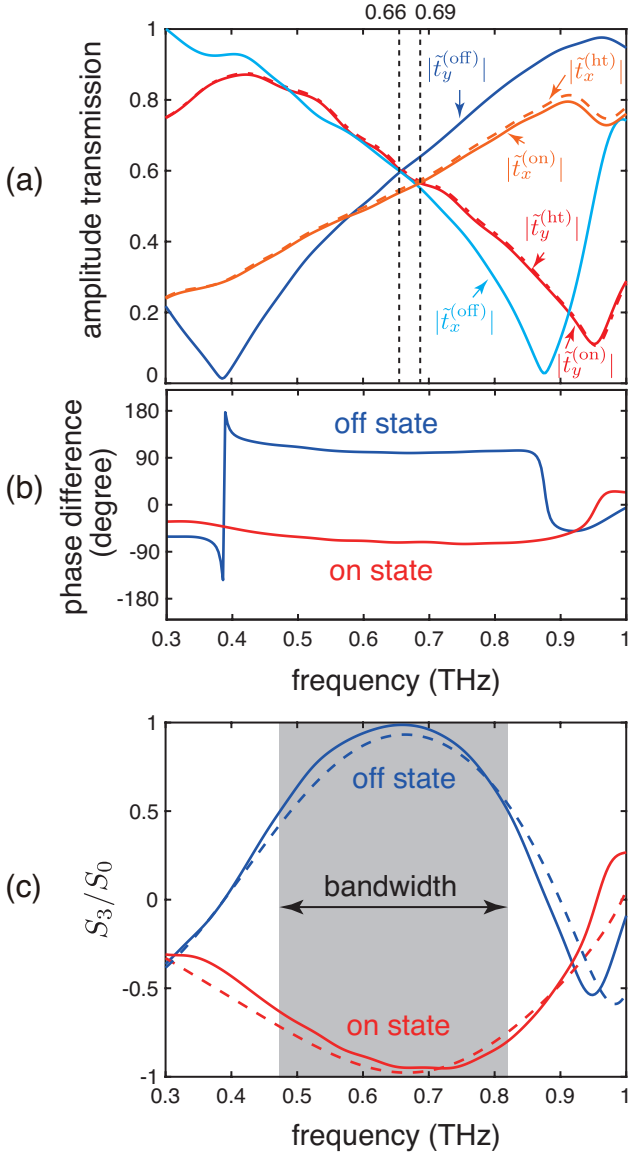}
  \caption{(a)
  Solid lines represent amplitude transmission spectra for
  off state at $I = 0$ and on state at $I = 180 \U{mA}$.
  Dashed line labeled as $\tilde{t}_x^{(\rm ht)}$ and
  dash-dotted line labeled as $\tilde{t}_y^{(\rm ht)}$
  represent the spectra at $85\U{^\circ C}$ for $x$ and
  $y$ polarizations, respectively. (b) Phase differences between $y$ and
  $x$ polarization components derived from experimental results for off state at
  $I = 0$ and on state at $I = 180\U{mA}$. (c) Normalized Stokes parameter
  $S_3/S_0$. Solid and dashed lines correspond to experimental and
  simulation results, respectively.
  Gray-shaded area corresponds to operation bandwidth satisfying
  $|S_3/S_0| > 0.5$ for both states.}
  \label{exp}
 \end{center}
\end{figure}

Subsequently, we evaluated the transmission characteristics of the metasurface
via conventional terahertz time-domain spectroscopy\cite{Hangyo2005},
in which complex transmission coefficients $\toff_x(\omega)$, $\toff_y(\omega)$, $\ton_x(\omega)$, and $\ton_y(\omega)$
are derived by the Fourier transformation of the obtained signals in the time domain.
The amplitude transmission is normalized by the reference signals,
which are obtained using the sapphire substrate without the metasurface. 
The derived magnitudes of the transmission coefficients are shown as solid lines in Fig.~\ref{exp}(a),
where $\toff_x$ and $\toff_y$ are obtained without current injection
and $\ton_x$ and $\ton_y$ are obtained at $I = 180 \U{mA}$.
The dashed and dash-dotted lines represent transmission spectra $\tilde{t}_x^{(\rm ht)}$ and  $\tilde{t}_y^{(\rm ht)}$
for the $x$ and $y$ polarizations, respectively,
 when the holder of the substrate 
is heated at $85\U{^\circ C}$ without current injection into the metasurface.
For both the polarization states, the obtained results at $I = 180 \U{mA}$
are almost the same as those for $85\U{^\circ C}$,
and it is evident that the current injection at $I = 180 \U{mA}$ is sufficient to induce a complete phase transition
in the \vo2 patches.
The four transmission spectra, $|\toff_x|$, $|\toff_y|$, $|\ton_x|$, and $|\ton_y|$, obtained in the experiment
agree well with the simulation results shown in Fig.~\ref{simulation2}(a).
Some discrepancy possibly arises from the experimental limitations, such as
wavefront deformation of the incident terahertz waves, 
fabrication error of the metasurface, and
finite conductivity of the aluminum films.
Figure~\ref{exp}(b) presents the phase differences between the $y$ and $x$ polarization components for both the states.
At $0.66 \U{THz}$ where $|\toff_x| = |\toff_y| = 0.60$ 
and at $0.69 \U{THz}$ where $|\ton_x| = |\ton_y| = 0.57$,
the phase differences are estimated as $+99\U{^\circ}$ and $-71\U{^\circ}$
for the off and on states, respectively.
The absolute power transmissions, including the Fresnel reflection loss of $-46\%$ at both sides of the substrate,
 are estimated as 20\,\% and 18\,\%, respectively.

To evaluate the function of the metasurface as an active quarter-wave plate,
we use $S_3/S_0 = 2 {\rm Im} (\tilde{t}_x^*
\tilde{t}_y)/(|\tilde{t}_x|^2+|\tilde{t}_y|^2)$,
which provides one of the normalized Stokes parameters for the incidence of a $45$-degree linear polarization\cite{Saleh}.
When the output wave is perfectly circularly polarized,
$S_3/S_0$ becomes $\pm 1$, 
whose sign corresponds to the helicity of the waves.
Figure \ref{exp}(c) presents the derived Stokes parameters for the off and on states.
The solid and dashed lines correspond to the experimental and simulation results, respectively.
It is confirmed that the helicity of the output terahertz wave
is reversed from $+0.99$ to $-0.95$ close to $f_0 = 0.66\,{\rm THz}$,
where the difference in $S_3/S_0$ is maximized,
and the fabricated metasurface acts as an active quarter-wave plate, as expected.
Defining the operation bandwidth, $\Delta f$, as a spectral
region satisfying $|S_3/S_0| > 0.5$ for both the states, represented by
the gray-shaded area in Fig.~\ref{exp}(c),
we estimated $\Delta f = 0.35 \U{THz}$ and the relative bandwidth, $\Delta f/f_0 = 0.52$.
The present study achieves a 4.2 times broader bandwidth
compared to 
a previous result, $\Delta f/f_0 = 0.12$, with a dipole-nested checkerboard metasurface,
which presents complex spectral shapes\cite{Nakata2018}.
Because the metasurface in this study is topologically equivalent to a
simple checkerboard structure
with a broad resonance,
we can achieve a flat phase response, as shown in Fig.~\ref{exp}(b),
which results in the broadband operation as an active quarter-wave plate.

In this study, we have demonstrated the broadband operation of a metasurface functioning as an
active quarter-wave plate, whose fast and slow axes can be interchanged.
Both the simulation and experimental results confirm that
the metasurface presents excellent performance as an active
quarter-wave plate,
and the available bandwidth is 4.2 times broader than that in a previous study.
The states of the metasurface are controlled by directly injecting
electric currents, which can also be utilized to monitor the electric
states of vanadium dioxide.
Compared with other related studies of electrically controllable metasurfaces with
${\rm VO_2}$ films, which are connected in parallel with metallic
elements \cite{Zhu2017, Han2017, Zhou2017, Nouman2018, Zhang2019},
the ${\rm VO_2}$ films of this metasurface are connected in series
in the current direction.
To substantially reduce the critical voltage for the series structures,
a supplementary heater formed of
titanium sheets is also integrated in the metasurface.
This method can be applicable to various types of active metasurfaces
employing vanadium dioxide.
The response time is estimated to be 60\,--90 seconds
 from the transient measurement of I--V characteristics for sudden current change.
Some studies have shown that W-doped \vo2 films have lower critical
temperature \cite{Horlin, Karaoglan-Bebek}, which could reduce the
transition time. Photoinduced phase transition by ultrafast optical
pulses \cite{Nakajima2008a, Xue2013a} might be the most effective way, which could reduce the transition time to picosecond order. 
The broadband active quarter-wave plate enables the polarization switching of
short terahertz pulses with a broad spectrum, which opens a new route for sensitive
detection of chiral molecules and terahertz data transmission.

See the Supplementary Material for the electric property of \vo2 film, the optimization of the design
parameters, and the fabrication procedures.

The metasurface was fabricated with the help of Kyoto University Nano
Technology Hub, as part of the ``Nanotechnology Platform Project''
sponsored by the MEXT in Japan.
The present research is supported by JSPS KAKENHI Grant Nos. 17K05075,
17K17777, and 20K05360, and the Shimadzu Science Foundation.

\noindent\\
{\sf DATA AVAILABILITY}\\
The data that support the findings of this study are available
from the corresponding author on a reasonable request.


%

\clearpage

\onecolumngrid
\begin{center}
{\large \bf Supplementary material:\\
Broadband operation of active terahertz quarter-wave plate achieved with vanadium-dioxide-based metasurface switchable by current injection}
\end{center}
\title[]{}


	     \maketitle


\twocolumngrid

 \section{Electric property of VO$_2$ film}    
\begin{figure}[b]
 \begin{center}
  \includegraphics[scale=0.95]{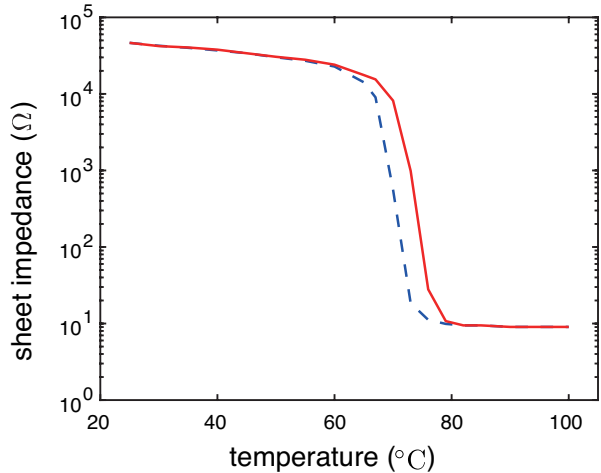}
  \caption{Typical sheet impedance of 200-nm-thick VO$_2$ film for various temperatures.
  The solid (dashed) line corresponds to the sheet impedance with increasing (decreasing) temperature. }
  \label{VO2}
 \end{center}
\end{figure}

Figure \ref{VO2} shows a typical sheet impedance of a 200-nm-thick \vo2 sheet,
which is fabricated by the procedures given in Sec.~III in the Supplementary Material,
for various temperatures.
This is obtained by a four-probe method, whose results can be applied for terahertz waves.
The solid (dashed) line corresponds to the sheet impedance with increasing (decreasing) temperature. 
The typical sheet impedances of 200-nm-thick \vo2 sheets are
$40\U{k\Omega}$\,--$150\U{k\Omega}$
at $25\U{^\circ C}$ and $8\U{\Omega}$\,--$12\U{\Omega}$ at $100\U{^\circ C}$, respectively.
Hence, we assume the sheet impedances as $10 \U{\Omega}$ for the metallic state and as $100\U{k\Omega}$
for the insulating state when designing the metasurface.

\section{Optimization of design parameters}

\begin{figure}[b]
 \begin{center}
  \includegraphics[]{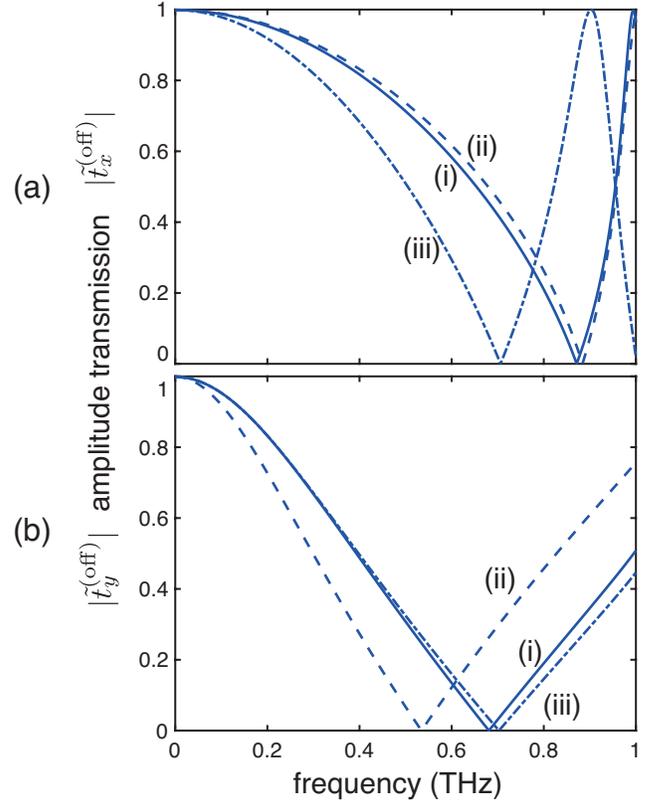}
  \caption{Normalized transmission spectra of off state for (a) $x$
  polarization and (b) $y$ polarization.
 Design parameters are (i) $b = 15 \U{\mu m}$, $c_1 = c_2 = 15 \U{\mu m}$;
 (ii) $b = 20 \U{\mu m}$, $c_1 = c_2 = 15 \U{\mu m}$;
 (iii) $b = 15 \U{\mu m}$, $c_1 = c_2 = 20 \U{\mu m}$. }
  \label{simulation}
 \end{center}
\end{figure}

Figures~\ref{simulation}(a) and (b) represent the normalized amplitude transmission spectra 
$|\toff_x|$ and $|\toff_y|$, respectively, for three sets of parameters:
(i) $b = 15 \U{\mu m}$, $c_1 = c_2 = 15 \U{\mu m}$ (solid lines);
(ii) $b = 20 \U{\mu m}$, $c_1 = c_2 = 15 \U{\mu m}$ (dashed lines);
(iii) $b = 15 \U{\mu m}$, $c_1 = c_2 = 20\U{\mu m}$ (dash-dotted lines).
The other parameters are fixed as $s = 10 \U{\mu m}$ and $a = 60 \U{\mu m}$.
The transmission spectra are normalized by that without the metasurface,
to exclude the Fresnel reflection at the surfaces of the sapphire substrate.
For the $x$ polarization, 
the spectra present significant red shifts with increasing $c_1$ (or $c_2$),
whereas they are insensitive to the change in $b$.
However, for the $y$ polarization,
the spectra present red shifts with increasing $b$,
whereas they are almost independent of $c_1$ (or $c_2$).
Consequently, it is possible to tailor the transmission spectra
for the $x$ and $y$ polarizations almost independently,
and we can easily adjust the dimensions of $b$, $c_1$, and $c_2$ to satisfy the conditions
to realize a quarter-wave plate, as expressed in  Eq.~(2).

\section{Fabrication procedures}

The metasurface is composed of three layers: a vanadium dioxide layer as 
a variable resistive sheet, an aluminum layer as a conductive sheet,
and a titanium layer as a supplemental heater.
The fabrication procedure of the metasurface is as follows.
A \vo2 film is deposited on a {\it c}-cut sapphire substrate of size $20 \U{
mm} \times 20 \U{mm} \times 1 \U{mm}$
by reactive magnetron sputtering with a vanadium target.
The thickness of the film is estimated as approximately $200 \U{nm}$.
After the \vo2 pattern is formed by wet etching,
metallic structures are patterned by a lift-off process
using a $400$-$\U{nm}$-thick aluminum film, which is deposited by electron-beam evaporation.
Finally, a supplemental heater is fabricated by a lift-off process using
a $24$-$\U{nm}$-thick titanium film formed by electron-beam evaporation.

\end{document}